\documentclass[%
 reprint,
superscriptaddress,
 amsmath,amssymb,
 aps,
]{revtex4-2}

\usepackage{graphicx}
\usepackage{dcolumn}
\usepackage{bm}
\usepackage{siunitx}
\usepackage[english]{babel}
\usepackage{hyperref}

\begin{document}


\title{SI-traceable frequency dissemination at 1572.06 nm in a stabilized fiber network with ring topology}
\author{Dominik Husmann}
\email{dominik.husmann@metas.ch}
\affiliation{Federal Institute of Metrology METAS, Lindenweg 50, 3003 Bern-Wabern, Switzerland}%
\author{Laurent-Guy Bernier}%
\affiliation{Federal Institute of Metrology METAS, Lindenweg 50, 3003 Bern-Wabern, Switzerland}
\author{Mathieu Bertrand}
\affiliation{ETH Zurich, Institute for Quantum Electronics, Auguste-Piccard-Hof 1, CH-8093 Zurich, Switzerland}
\author{Davide Calonico}
\affiliation{INRIM Istituto Nazionale di Ricerca Metrologica, I-10135 Torino, Italy}
\author{Konstantinos Chaloulos}
\affiliation{SWITCH, Werdstrasse 2, 8021 Zurich, Switzerland}
\author{Gloria Clausen}
\affiliation{ETH Zurich, Laboratory of Physical Chemistry, CH-8093 Zurich, Switzerland}
\author{Cecilia Clivati}
\affiliation{INRIM Istituto Nazionale di Ricerca Metrologica, I-10135 Torino, Italy}
\author{Jérôme Faist}
\affiliation{ETH Zurich, Institute for Quantum Electronics, Auguste-Piccard-Hof 1, CH-8093 Zurich, Switzerland}
\author{Ernst Heiri}
\affiliation{SWITCH, Werdstrasse 2, 8021 Zurich, Switzerland}
\author{Urs Hollenstein}
\affiliation{ETH Zurich, Laboratory of Physical Chemistry, CH-8093 Zurich, Switzerland}
\author{Anatoly Johnson}
\affiliation{Department of Chemistry, University of Basel, Klingelbergstrasse 80, 4056 Basel, Switzerland}
\author{Fabian Mauchle}
\affiliation{SWITCH, Werdstrasse 2, 8021 Zurich, Switzerland}
\author{Ziv Meir}
\affiliation{Department of Chemistry, University of Basel, Klingelbergstrasse 80, 4056 Basel, Switzerland}
\author{Frédéric Merkt}
\affiliation{ETH Zurich, Laboratory of Physical Chemistry, CH-8093 Zurich, Switzerland}
\author{Alberto Mura}
\affiliation{INRIM Istituto Nazionale di Ricerca Metrologica, I-10135 Torino, Italy}
\author{Giacomo Scalari}
\affiliation{ETH Zurich, Institute for Quantum Electronics, Auguste-Piccard-Hof 1, CH-8093 Zurich, Switzerland}
\author{Simon Scheidegger}
\affiliation{ETH Zurich, Laboratory of Physical Chemistry, CH-8093 Zurich, Switzerland}
\author{Hansjürg Schmutz}
\affiliation{ETH Zurich, Laboratory of Physical Chemistry, CH-8093 Zurich, Switzerland}
\author{Mudit Sinhal}
\affiliation{Department of Chemistry, University of Basel, Klingelbergstrasse 80, 4056 Basel, Switzerland}
\author{Stefan Willitsch}
\affiliation{Department of Chemistry, University of Basel, Klingelbergstrasse 80, 4056 Basel, Switzerland}
\author{Jacques Morel}%
\affiliation{Federal Institute of Metrology METAS, Lindenweg 50, 3003 Bern-Wabern, Switzerland}%

\date{\today}

\begin{abstract}
Frequency dissemination in phase-stabilized optical fiber networks for metrological frequency comparisons and precision measurements are promising candidates to overcome the limitations imposed by satellite techniques. However, network constraints restrict the availability of dedicated channels in the commonly-used \textit{C-band}. Here, we demonstrate the dissemination of an SI-traceable ultrastable optical frequency in the \textit{L-band} over a \SI{456}{km} fiber network with ring topology, in which telecommunication data traffic occupies the full \textit{C-band}. We characterize the optical phase noise and evaluate a link instability of \SI{4.7e-16} at \SI{1}{s} and \SI{3.8e-19} at \SI{2000}{s} integration time, and a link accuracy of \SI{2e-18}{}, which is comparable to existing metrology networks in the \textit{C-band}. We demonstrate the application of the disseminated frequency by establishing the SI-traceability of a laser in a remote laboratory. Finally, we show that our metrological frequency does not interfere with data traffic in the telecommunication channels. Our approach combines an unconventional spectral choice in the telecommunication \textit{L-band} with established frequency-stabilization techniques, providing a novel, cost-effective solution for ultrastable frequency-comparison and dissemination, and may contribute to a foundation of a world-wide metrological network.

\end{abstract}

\maketitle


\section{Introduction}
Precise dissemination of accurate frequency signals traceable to the SI definition of the second is essential in many scientific fields, such as precision spectroscopy \cite{matveev_precision_2013,safronova18a,clivati_measuring_2016,santagata_high-precision_2019}, remote clock comparisons in fundamental metrology \cite{lisdat_clock_2016,delva_test_2017,hong_measuring_2009,beloy_frequency_2021}, relativistic geodesy \cite{takano_geopotential_2016,grotti_geodesy_2018}, or synchronization in large-scale facilities \cite{clivati_common-clock_2020,he_long-distance_2018,krehlik_fibre-optic_2017}. In particular, a redefinition of the SI second, as is currently under evaluation, necessitates the comparison of state-of-the-art optical clocks with $10^{-18}$ relative frequency uncertainty \cite{riehle_cipm_2018}. Most of these applications require higher resolution than allowed by established satellite techniques based on two-way time and frequency transfer (TWTFT) or global navigation satellite systems (GNSS), which achieve fractional frequency stabilities at the $10^{-16}$ level with measurement times of a few days \cite{petit_1_2015}.


\begin{figure*}[ht]
	\centering\includegraphics[width=0.8\textwidth]{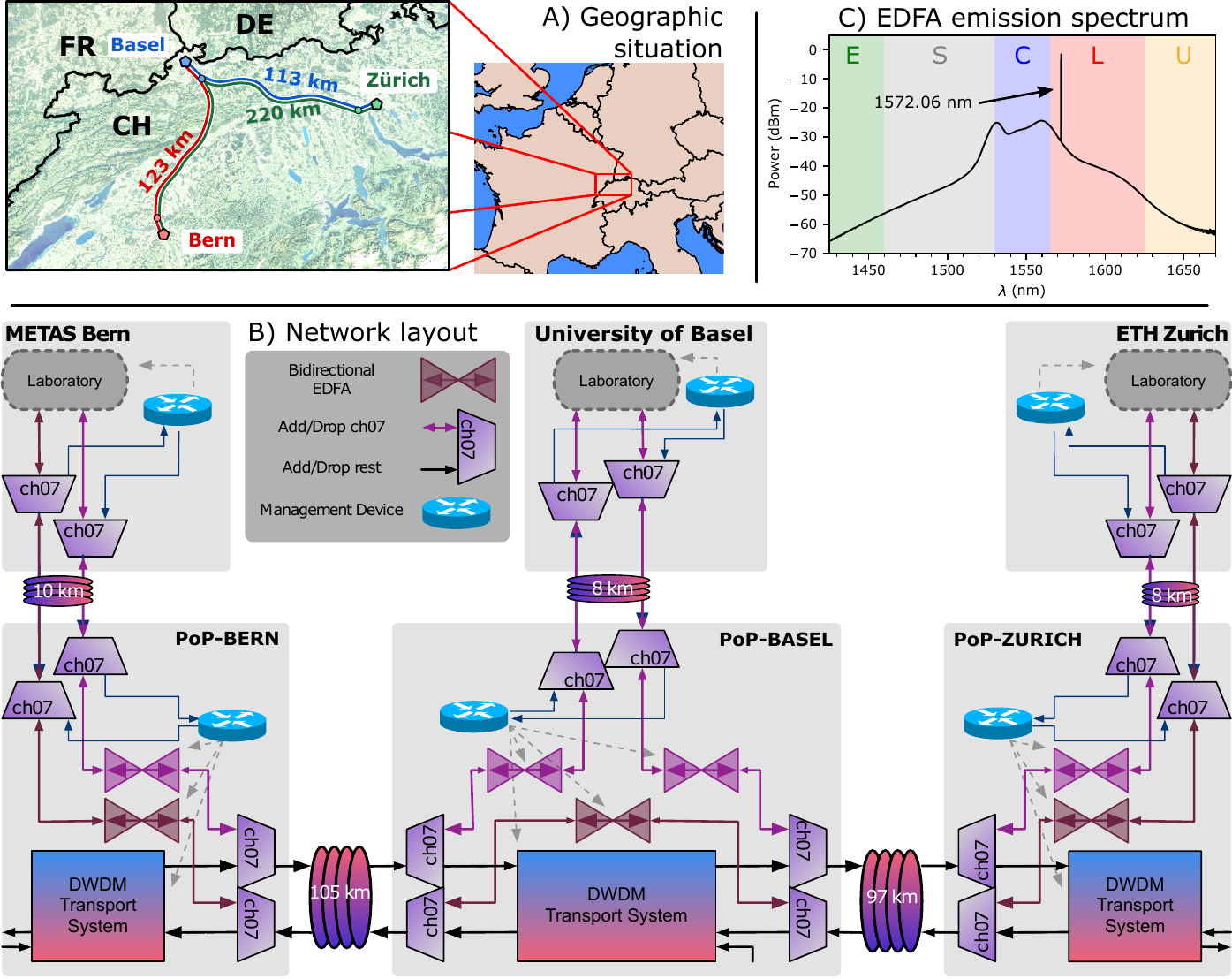}
	\caption{A) Geographical situation of the Swiss metrology network connecting institutes in Bern, Basel and Zurich. Each network leg (red, blue and green lines) connects two institutes (big pentagons), and passes through 2-3 network points of presence (PoPs, small circles), where the signal is amplified. B) Emission spectrum of the bidirectional erbium-doped fiber amplifier, when injecting a laser at \SI{1572}{nm} with a power of around \SI{-20}{dBm},  measured with a resolution bandwidth of \SI{0.2}{nm}. The letters indicate the spectral bands according to ITU-T recommendation. C) Integration of the metrological channel CH07 into the data network. Purple and brown arrows indicate the bidirectional metrological signal, black arrows indicate the unidirectional data traffic. Bidirectional EDFAs are used to amplify the metrological signal, and \SI{100}{GHz} wide OADM (ports Add/Drop CH07, Add/Drop rest) filters allow inserting and ejecting the CH07 signal from the SWITCH backbone in the PoPs. Management devices (routers and switches) enable remote control over network and laboratory components (blue arrows).}
	\label{fig1}
\end{figure*}

To overcome the limitations of satellite-based frequency comparison techniques, several phase-stabilized optical fiber networks for the dissemination of ultra-stable and accurate optical frequencies have been implemented recently \cite{matveev_precision_2013,clivati_measuring_2016,santagata_high-precision_2019,beloy_frequency_2021,guena_first_2017,lisdat_clock_2016,delva_test_2017,hong_measuring_2009,grotti_geodesy_2018,droste_optical-frequency_2013,williams_high-stability_2008,lopez_frequency_2015,predehl_920-kilometer_2012,clivati_common-clock_2020,fujieda_all-optical_2011,he_long-distance_2018,krehlik_fibre-optic_2017}, spanning thousands of kilometers and providing transfer stabilities of few $ 10^{-15} $ at \SI{1}{s} with ultimate accuracies beyond $10^{-19}$ \cite{cantin_accurate_2021}. The majority of these networks operate in dedicated fibers (\textit{dark fibers}) and are spectrally situated in the International Telecommunication Union Channel 44 (ITU-T CH44, wavelength $\lambda=\SI{1542.14}{nm}$) in the \textit{C-band} (\SIrange{1530}{1565}{nm}), where the optical loss is minimal and off-the-shelf telecommunication components are available \cite{union_optical_2010}. However, the high recurring costs for dark-fiber lease have hindered a wider development of such networks. Sharing the available spectrum of the fiber with other network users by exploiting the dense wavelength division multiplexing (DWDM) architecture can significantly reduce costs. A fixed \SI{100}{GHz} dark channel in the \textit{C-band} has been implemented in France \cite{cantin_accurate_2021} to integrate the ultra-stable frequency signal within the telecommunications data network. However, network operators may have concerns with such a choice, as state-of-the-art modulation techniques rely on flexible spectral usage based on re-configurable add-drop multiplexers (ROADM), which conflicts with the presence of such a fixed alien channel inside the \textit{C-band} that might limit or block full spectral use and future upgrades. Furthermore, optical networks often incorporate highly integrated amplification structures, which do not a priori account for accommodation of alien signals in the \textit{C-band}. The growing network occupancy of the \textit{C-band} and the restrictions imposed by modern optical fiber network infrastructure thus limit the availability of dedicated metrology channels and leading to higher rental costs. Hence, alternatives in less densely occupied spectral bands need to be explored and validated.

Here, we report the realization of a stabilized frequency-metrology network in Switzerland (see Fig.~\ref{fig1}A), spanning over \SI{456}{km} of optical fibers, and operating in the \textit{L-band} (\SIrange{1565}{1625}{nm}, see Fig.~\ref{fig1}C) in ITU-T CH07 (\SI{1572.06}{nm}). The metrological frequency channel allows the dissemination of an ultrastable SI-traceable frequency between dedicated research institutes. The metrological signal is hosted in the same fiber as the data traffic of the Swiss National Research and Education Network (NREN) operated by SWITCH, which lies entirely in the \textit{C-band}. The choice of ITU-T CH07 considers both the advantage provided by a channel outside the \textit{C-band} as discussed above, and the availability and performance of optical components such as narrow-linewidth diode laser, erbium-doped fiber amplifiers (EDFAs) and optical add-drop multiplexers (OADM). Moreover, the spectral separation minimizes the risk of interference to and from the densely used data-traffic channels, which are separated by spectral guard bands of several nanometers. As the optical components were by default not commercially available in the \textit{L-band} at the start of this project, the components used here were specifically tailored to our wavelength by the manufacturers. The growing need for \textit{L-band} components, however, has in the meantime led to an increase in their commercial availability.

The metrology network connects the Swiss Federal Institute of Metrology, METAS, where an SI-traceable optical frequency is generated, with the University of Basel and ETH Zurich, where the disseminated frequency provides a new improved reference for precision spectroscopy laboratories \cite{sinhal_quantum-nondemolition_2020,germann_observation_2014,beyer_metrology_2018,peper_precision_2019}. The network consists of three individually phase-stabilized fiber legs laid out in a ring topology (see Fig.~\ref{fig1}A), which allows for a genuine end-to-end evaluation of the stability and accuracy of the disseminated frequency.
We validate this approach by demonstrating an end-to-end stability of \SI{4.7e-16} at \SI{1}{s} and \SI{3.8e-19} at \SI{2000}{s}, and a link accuracy of \SI{2e-18}{} in the disseminated optical frequency over \SI{456}{km}. 
As a first application, we show an improvement of the traceability of a \SI{729}{nm} laser used in precision spectroscopy in one of the remote stations. The level of improvement renders novel regimes of precision measurements in atomic and molecular physics accessible, facilitating ways to address fundamental physics questions such as those related to the proton-radius puzzle, a possible variation in time of fundamental constants, and possible extensions of the standard model of particle physics \cite{safronova18a,matveev_precision_2013}.

\begin{figure*}[ht]
	\centering\includegraphics[width=0.8\textwidth]{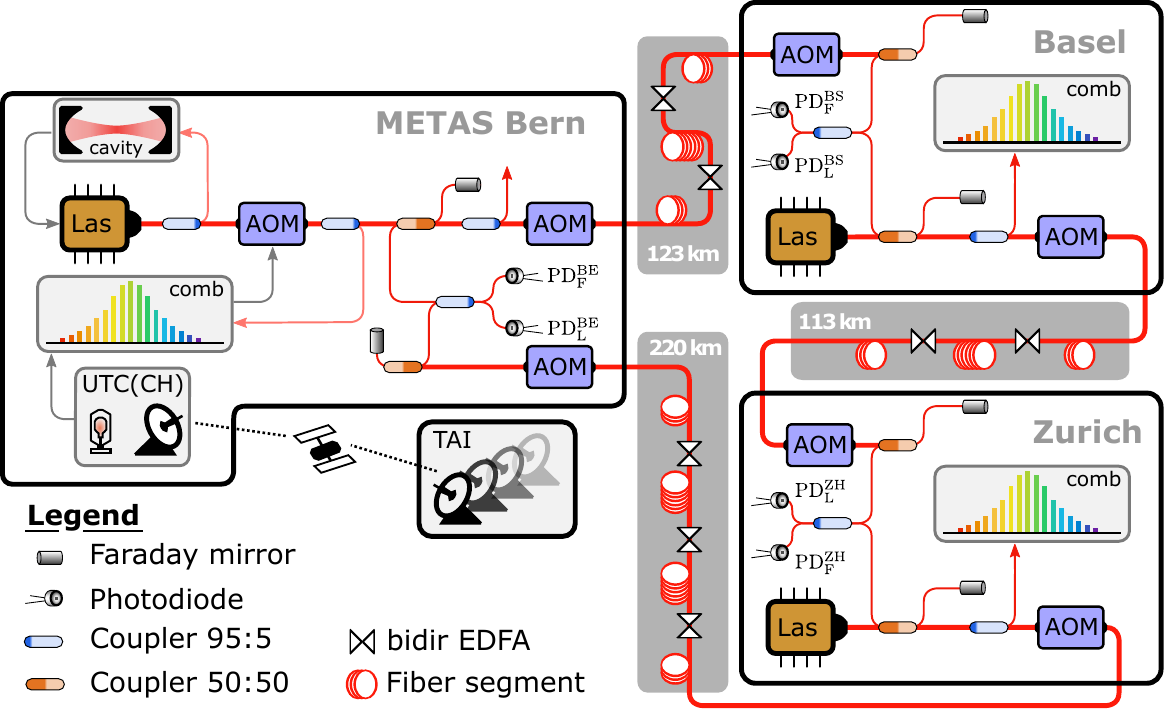}
	\caption{Detailed schematic of the optical layout: Master station METAS Bern (BE) with traceability to TAI via UTC(CH), regeneration stations at University Basel (BS) and ETH Zurich (ZH), and phase-noise cancellation (PNC) systems with corresponding acousto-optic modulators (AOMs). Optical attenuation in the three fiber legs (gray areas) is compensated using bidirectional EDFAs. Two photodiodes per station detect the fiber PNC beat (F) and laser beat (L) signals. Each station hosts a laser (Las) at \SI{1572.06}{nm} and an optical frequency comb (comb). Additional electronic filtering components are omitted in the figure for clarity.}
	\label{fig2}
\end{figure*}

\section{Experimental aspects}
\subsection{Network layout}
The three optical fiber legs forming the network (see Tab.~\ref{tab:network} and Fig.~\ref{fig1}A) connect METAS in Bern (BE) to the University of Basel (BS), the University of Basel to ETH Zurich (ZH) and ETH Zurich back to METAS, where the last fiber leg geographically follows the first two fiber segments (see Fig.~\ref{fig1}A). 
The advantage of such a ring network topology lies in the capability for the direct evaluation of the end-to-end optical frequency dissemination uncertainty. Each leg is individually stabilized with the well established phase-noise cancellation (PNC) scheme \cite{ma_delivering_1994}. In the master station at METAS, the metrological optical frequency of \SI{190.7}{THz} (wavelength \SI{1572.06}{nm}) is prepared and referenced to Coordinated Universal Time (UTC). In the laboratories in Basel and Zurich, the frequency signal is regenerated and sent downstream while part of it is coupled out for local use.
The fibers are part of the Swiss NREN (SWITCH). The network implementation is shown in detail in Fig.~\ref{fig1}B. We use DWDM with \SI{100}{GHz} OADMs to insert and extract the metrological signal with frequency in ITU-T CH07 into the network. Each leg consists of a long-haul intercity segment, and two shorter urban segments connecting the institutes to point-of-presence (PoP) backbone nodes of the SWITCH network. Power losses in the fibers and in optical add-drop multiplexer modules are compensated using bidirectional EDFAs (\textit{Czech Optical Solutions, CLA BiDi}) specifically tailored to \SI{1572}{nm} (see Fig.~\ref{fig1}C). These amplifiers are sandwiched between two OADM filters to suppress amplified spontaneous emission (ASE) outside our \SI{100}{GHz} channel and to limit self-lasing from amplified reflections \cite{vojtech_white_2020}. The level of amplification was set for each EDFA individually, with typical values of around \SIrange{15}{23}{dB}.  Optimally, the input power and amplification for the two directions of a bidirectional EDFAs should be kept symmetric. In our network, the placement of the EDFAs was restricted to the PoPs shown in Fig.~\ref{fig1}A, which did not always allow to satisfy this condition. However we did not encounter any problems arising from this asymmetry. Ultimately, we chose the amplification setting in a way that the optical power launched into the fiber never exceeded $\approx$\SI{6}{dBm} in order to reduce detrimental effects from Brillouin and Rayleigh scattering.

\begin{table*}
	\centering
	\begin{tabular}{c|c|c|c|c}
		Leg & Length (km) & 1-way Att. (dB) & \# EDFAs & delay limit (Hz)\\ \hline\hline
		METAS - Uni Basel & 123 & 31& 2 & 406\\
		Uni Basel - ETH Zurich & 113 & 29 & 2& 442\\
		ETH Zurich - METAS & 220 & 51 & 3 & 227
	\end{tabular}
	\caption{Overview of the three legs of the stabilized frequency-dissemination network. Listed are the length of each leg, the 1-way attenuation measured in the \textit{C-band}, the number of EDFAs, and the delay limit frequency.}
	\label{tab:network}
\end{table*}

\subsection{Master and regeneration lasers}
The main components of the stabilized laser, the PNC systems, and the regeneration stations at the three laboratories are depicted in  Fig.~\ref{fig2}. The regeneration stations and PNC systems presented here were designed and built by Istituto Nazionale di Ricerca Metrologica (INRIM). The stable frequency at METAS is generated by locking an external-cavity diode laser (\textit{RIO Planex}) at a wavelength of \SI{1572.06}{nm} to an ultra-low expansion (ULE) cavity 
of finesse 13'000 using a Pound-Drever-Hall locking scheme. This laser acts as the master and its frequency is continuously measured against an SI-referenced frequency comb (\textit{FC1500-250-ULN, Menlo Systems}). Slow drifts in the laser frequency due to residual thermal relaxation of the cavity are compensated using a digital feedback loop acting on an acousto-optic modulator (AOM). The optical frequency comb is referenced by an active maser, which contributes to the realization of the Swiss timescale UTC(CH) and provides a short-term stability of $\SI{1.7e-13}{}$ at \SI{1}{s}. The phase of the maser is regularly compared to the International Atomic Time (TAI) via UTC(CH). This comparison allows us to determine the drift of the maser and thus to estimate its absolute frequency traceable to the SI definition of the second at all times.
The stable and traceable optical frequency is then fed into a phase-stabilized link to the next terminal in Basel. The offset-locked regeneration lasers at the University of Basel and ETH Zurich are identical: a local external cavity diode laser (\textit{RIO Planex}) with a free-running linewidth of around \SI{2}{kHz} is phase locked to the incoming optical signal with an offset frequency of \SI{-60}{MHz}, given by the beat signal detected on photodiodes PD$_{\mathrm{L}}^{\mathrm{BS}}$ and PD$_{\mathrm{L}}^{\mathrm{ZH}}$. We limit the bandwidth of this offset lock to few tens of kilohertz and thus regenerate the optical signal by discarding noise contributions caused by residual ASE of the EDFAs. A part of the signal from the regeneration laser is split out for local use (e.g. locking an optical frequency comb, see Sec. \ref{sec:729} below), while the remainder is injected into the subsequent PNC system and sent further downstream.

\subsection{Phase-noise cancellation}
Each of the three stations hosts a PNC system that stabilizes the subsequent fiber leg between the local and a remote station. The PNC system consists of a first AOM (\SI{35}{MHz}) at the local end that applies a frequency correction to the outgoing signal. At the remote end, a second AOM shifts the incoming signal by a fixed frequency of \SI{45}{MHz}. A coupler with a Faraday mirror reflects part of the signal back to the local end (see Fig.~\ref{fig2}). 
The second AOM allows one to distinguish the full two-way round-trip signal from detrimental reflections. The AOM frequencies are chosen to be incommensurate with the PNC full-round-trip beat note at $2\times(\SI{35}{MHz}+\SI{45}{MHz})=\SI{160}{MHz}$ in order to prevent amplitude modulation from spurious reflections in the fiber link. Note that the net frequency shift between two consecutive stations in the network, given by the AOM frequencies and the offset lock frequency, is $\SI{35}{MHz}+\SI{45}{MHz}-\SI{60}{MHz}=\SI{+20}{MHz}$, and the frequency shift of the signal that travels around the full loop is $3\times(\SI{35}{MHz}+\SI{45}{MHz})-2\times\SI{60}{MHz}=\SI{120}{MHz}$.

The round-trip signal of each of the PNC arms is overlapped with a local copy of the original signal in an all-fiber Michelson-type interferometer, and the beat signal is detected with the photodiodes PD$_{\mathrm{F}}^{\mathrm{BE}}$, PD$_{\mathrm{F}}^{\mathrm{BS}}$ and PD$_{\mathrm{F}}^{\mathrm{ZH}}$. Two tracking voltage-controlled oscillators (VCO) at each station are locked to this beat signal by phase-locked loops (PLL) allowing noise rejection outside the locking bandwidth and thus cleaning of the signal. The phase of the primary VCO is continuously compared to an RF reference to extract the fiber phase noise. A proportional-integral (PI) feedback circuit applies a proper frequency correction to the optical signal via the local AOM that stabilizes its frequency against the detected perturbations. The PI parameters are matched to the fiber length by ensuring the \SI{0}{dB} gain to be at a frequency below the delay limit (see Sec.~\ref{sec:phasenoise}). The second auxiliary VCO has a slightly different locking bandwidth and is used as a redundant counter that enables to spot cycle slips (CS) \cite{udem_accuracy_1998} in the PNC.

\section{Results and discussion}
\subsection{Phase-noise measurement}
\label{sec:phasenoise}
\begin{figure*}[ht]
	\centering\includegraphics[width=1\textwidth]{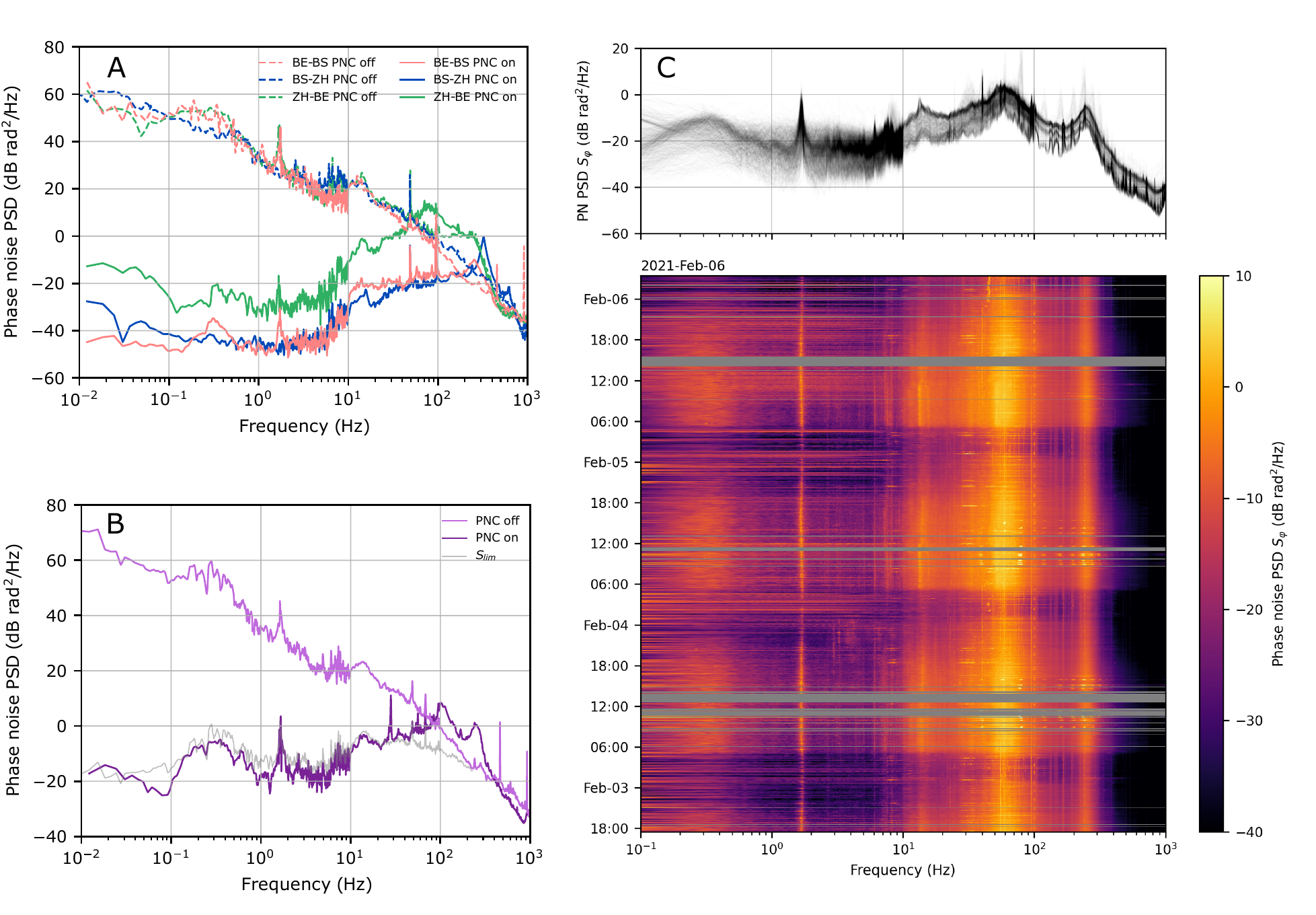}
	\caption{A) Phase noise  PSD of the \SI{160}{MHz} round-trip PNC beat note of the three fiber segments  BE-BS (red), BS-ZH (blue) and ZH-BE (green), with PNC on (solid lines) and off (dashed lines). B) Phase noise PSD of the \SI{120}{MHz} end-to-end beat note between the local laser at METAS and the return signal from Zurich, with PNC off (light magenta) and on (dark magenta). Gray curve: theoretical limit imposed by delay-unsuppressed noise of the longest fiber leg ZH-BE. C) Same signal as in B in stabilized condition, measured over several days in February 2021 in intervals of around \SI{6}{min}. The upper panel shows all measurements superimposed onto each other. The lower panel shows a 2D view of the same data, with the y-axis representing time. Day-night cycles are distinctly visible as alternating high- and low-noise periods. Traces with excessive CS leading to high phase noise at low frequencies are omitted from the upper panel and masked in gray in the lower panel.}
	\label{fig3}
\end{figure*}

We characterized the phase-noise cancellation on the three individual fiber legs by measuring the in-loop phase-noise power spectral density (PSD) of their respective \SI{160}{MHz} beat signals detected on PD$_{\mathrm{F}}^{\mathrm{BE}}$, PD$_{\mathrm{F}}^{\mathrm{BS}}$ and  PD$_{\mathrm{F}}^{\mathrm{ZH}}$. The results are presented in Fig.~\ref{fig3}A. We see qualitative congruence between the three fiber legs. In absence of PNC, the noise spectrum is dominated by white frequency noise. We observe an increased noise around \SI{10}{Hz}, similar to what was observed in other long-haul fiber links \cite{calonico_high-accuracy_2014,droste_optical-frequency_2013} where it had been attributed to frequencies originating from buildings and infrastructure. The PNC efficiently suppresses noise at frequencies below the delay limit of $\frac{1}{4\tau}$ \cite{williams_high-stability_2008} where $\tau$ is the one-way delay time of the signal passing through the fiber. In the case of our network, this delay limit amounts to \SI{406}{Hz}, \SI{442}{Hz} and \SI{227}{Hz} for the three network legs (see Tab.~\ref{tab:network}). The feedback loop gain for each fiber segment was chosen as a trade-off between an efficient low-frequency noise suppression and increased noise peaks around the delay-limited servo bandwidth.

Though the in-loop phase-noise measurement gives an indication of the overall fiber noise and the spectral response of the PNC, it carries insufficient information about the actual phase noise on the receiver end. In order to perform an end-to-end comparison of the frequency stability, we exploited the ring topology of our network and measured the phase noise of the return signal with respect to the original signal in Bern by detecting their beat on PD$_{\mathrm{L}}^{\mathrm{BE}}$. The resulting phase noise PSD of this \SI{120}{MHz} beat signal is shown in Fig.~\ref{fig3}B. Analogous to the results in Fig.~\ref{fig3}A, noise is efficiently rejected for frequencies below around \SI{100}{Hz}, which is slightly below the delay limit of the longest leg (ZH-BE). In spectral regions dominated by white frequency noise, this leads to a flat curve limited by delay-unsuppressed residual phase noise. We compare this to the theoretical limit of ideal noise rejection $S_{\rm lim} (f)$ \cite{williams_high-stability_2008} (gray curve in Fig.~\ref{fig3}B) for the longest fiber segment of \SI{220}{km}, under the assumption that noise is uncorrelated with position. The theoretical limit matches the measurement for frequencies below the servo bump, confirming that the noise suppression works at its optimum. We consider the phase noise in Fig.~\ref{fig3}B to be an upper bound for the phase noise of individual segments.

The fibers used in this project are typically installed close to main traffic routes, and are thus expected to be subjected to temporal fluctuations in environmental anthropogenic noise. We analyzed this by measuring the phase noise of the end-to-end beat over several days, in intervals of roughly \SI{6}{min}. The results are presented in Fig.~\ref{fig3}C and show distinct day-night oscillations: At around 5 am, there is a sharp increase in phase noise extending over the full spectrum. In the late afternoon we observe a gradual decrease in phase noise, leading to a minimum during the night. We attribute this evolution to environmental noise given by human activity and transport-system behavior reflecting typical work-day schedules. The distinct peak in phase noise at \SI{1.7}{Hz} originates from a fiber segment between Bern and Basel (see also Fig.~\ref{fig3}A). While its origin is not known, we consider resonances of particular infrastructures such as bridges or rotating fans in power plants located in the vicinity of the fiber as possible sources.

\subsection{Long-term stability of the link}
\label{sec:alldev}
\begin{figure}[ht]
	\centering\includegraphics[width=0.48\textwidth]{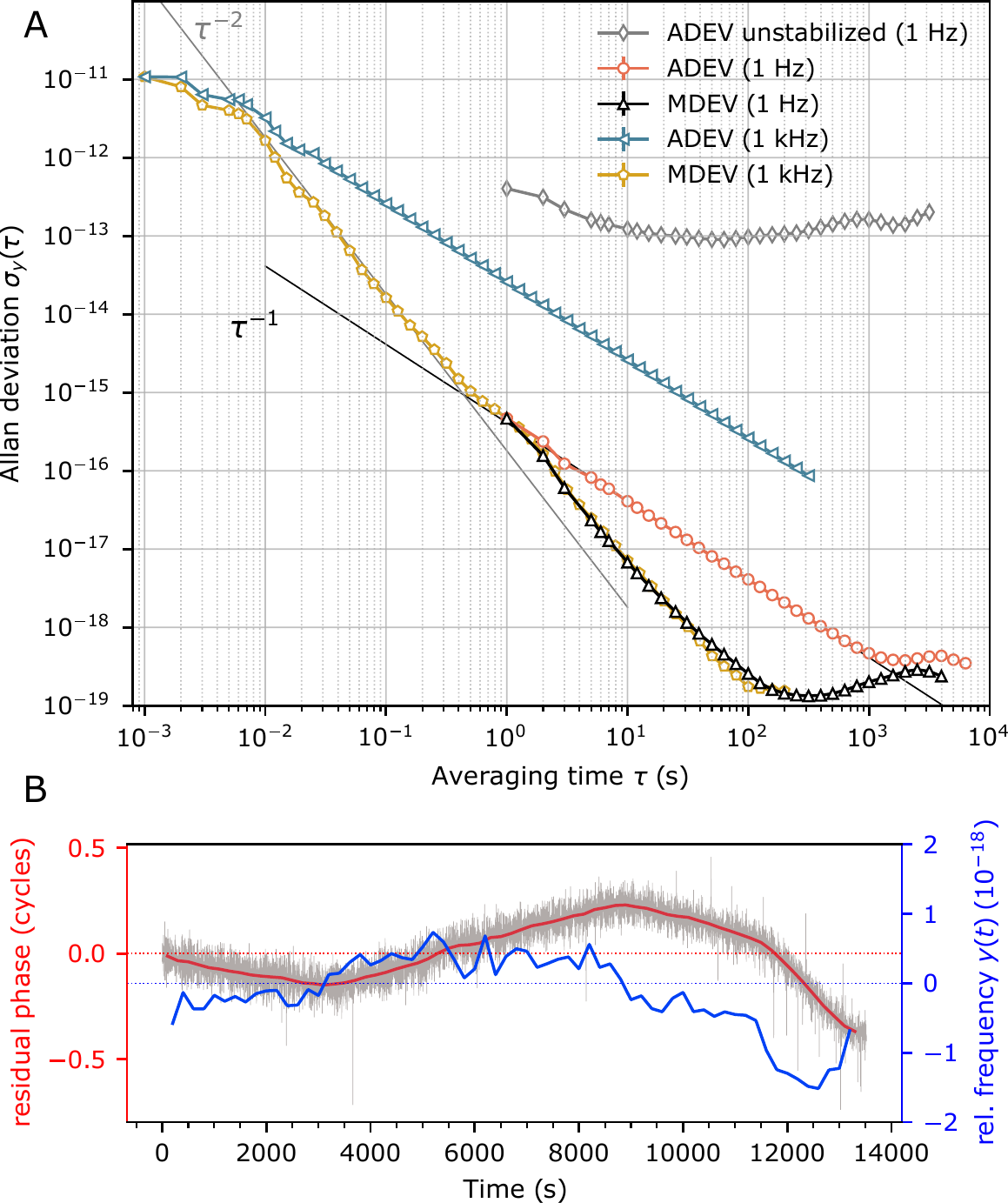}
	\caption{A) Allan deviation of the end-to-end beat frequency of the stabilized link. Red: ADEV with \SI{1}{Hz} sampling in $\Lambda-$type averaging mode. Black: MDEV of the same data. Blue: ADEV with \SI{1}{kHz} sampling. Yellow: MDEV with \SI{1}{kHz} sampling.	Gray: ADEV of the unstabilized link with \SI{1}{Hz} sampling. The gray straight line is a $1/\tau^2$ fit to a subset of the yellow data points. The black straight line is a $1/\tau$ fit to a subset of the red data points.
	B) Gray: Residual phase evolution of the \SI{1}{Hz} sampled data set from A. Red: moving average of the grey curve, with averaging window length of \SI{200}{s}. Blue: derivative of the red curve, corresponding to the deviation of the relative frequency from the nominal frequency. The magnitude of these excursions give an estimate of the frequency uncertainty introduced by the link.}
	\label{fig4}
\end{figure}

We analyzed the long-term stability of the fiber link by monitoring the full-loop end-to-end beat frequency on PD$_{\rm L}^{\rm BE}$ using a dead-time free counter (\textit{K+K FXE80}) \cite{kramer_extra_2004}. In a first set of experiments, we measured for around \SI{700}{s}  with a sampling rate of \SI{1}{kHz}. In a second set of measurements, we monitored the beat frequency over several hours, during which we selected the internal $\Lambda-$averaging mode of the frequency counter to reduce the sampling rate from \SI{1}{kHz} to \SI{1}{Hz} and to reduce the equivalent measurement bandwidth. For comparison, we repeated this measurement for the unstabilized link. The results are presented in Fig.~\ref{fig4}A, and show a link instability of \SI{4.7e-16}{} at \SI{1}{s} measurement time. The overlapping Allan deviation (ADEV) of the stabilized link shows a $1/\tau$ dependence, which is a typical behavior for fiber links affected by phase noise in the acoustic range \cite{newbury_coherent_2007,droste_optical-frequency_2013,calosso_avoiding_2016}, and it reaches a noise floor of \SI{3.8e-19} at \SI{2000}{s}. Owing to a stronger filtering of fast noise processes \cite{calosso_avoiding_2016,droste_optical-frequency_2013}, the modified Allan deviation (MDEV) shows a stronger decay of $1/\tau^2$, reaching a stability floor of \SI{1.3e-19} in the form of a local minimum at \SI{300}{s}. The MDEVs at \SI{1}{kHz} and \SI{1}{Hz} can be concatenated at $\tau=\SI{1}{s}$, in agreement with the fact that the low-pass filtering of the $\Lambda-$type averaging of the frequency counter is equivalent to the filtering performed in the calculation of the MDEV \cite{calosso_avoiding_2016}.

The stability analysis does not account for systematic shifts with respect to the nominal frequency, which may occur due to drifts in the uncompensated interferometer arms. To estimate these systematic deviations, we measured the relative frequency deviation from the nominal value for a CS-free segment measured over \SI{13500}{s} (see Fig.~\ref{fig4}B). We attribute these excursions to daily temperature variations in the uncompensated interferometer arms. The typical frequency excursions are of the order of $\SI{1e-18}{}$, based on which we determine a conservative limit of the link accuracy to be \SI{2e-18}{} under the assumption that residual interferometer noise in the individual stations is uncorrelated. For the application of frequency dissemination to remote stations, the current achieved link stability is two orders of magnitude better than the disseminated frequency, and thus does not represent a limitation. Nevertheless, for further upgrades, the link stability and accuracy could be improved by a refined design of the interferometer, e.g., designing a fully balanced interferometer \cite{cantin_accurate_2021,stefani_tackling_2015}.

\subsection{SI-referencing a \SI{729}{nm} spectroscopy laser at a remote station}
\label{sec:729}
\begin{figure}[ht]
	\centering\includegraphics[width=0.45\textwidth]{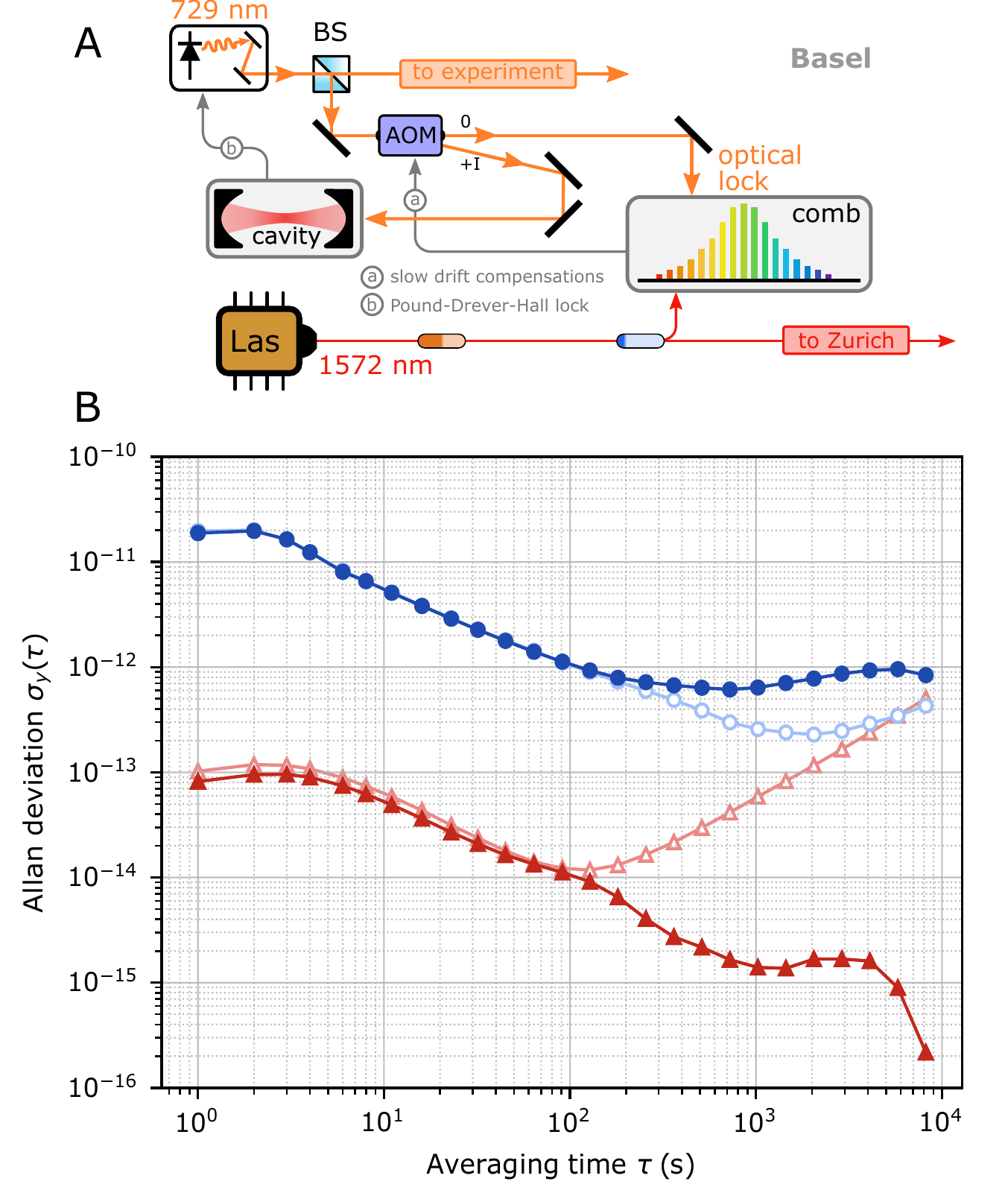}
	\caption{A) Compensation setup for improving the long-term stability of a \SI{729}{nm} spectroscopy laser in the remote station in Basel. B) Allan deviation of the frequency of a spectroscopy laser (\SI{729}{nm}) in the Basel remote station determined by two methods. For the  blue circles, the laser frequency is referenced to a local GPSD Rb standard (Eq.~\ref{eq:Rb}). For the red triangles, the laser frequency is referenced to the METAS standard via the fiber link (Eq.~\ref{eq:METAS}). The light-blue and light-red traces were obtained after the compensation for the slow drifts of the ULE cavity in Basel was turned off. Differences between the light and thick red traces at short times ($<\SI{100}{s}$) and the light and thick blue traces at long times ($>\SI{100}{s}$) are due to non-stationary effects, such as changes in the gain of the laser locks due to changes in the optical power levels.}
	\label{fig5}
\end{figure}

As a demonstration of the application of the frequency-dissemination network, we establish SI-traceabiltiy of the frequency of a narrow-linewidth near-infrared spectroscopy laser at \SI{729}{nm} in the remote station at the University of Basel by referencing it to the frequency standard at METAS (see Fig.~\ref{fig5}A for the locking setup). The short-term linewidth of the \SI{729}{nm} laser in Basel was stabilized by a lock to a local high-finesse ULE cavity. 
Without further corrections, the long-term stability was limited by the drifts of the ULE cavity of $\sim\SI{15}{mHz/s}$ (light-red trace in Fig.~\ref{fig5}B). We used the beat signal of the referenced regeneration laser with the local comb at \SI{1572.06}{nm} to track the slow drifts of the ULE cavity and subsequently compensate for them via a digital feedback loop acting on an AOM in the spectroscopy-laser path to the cavity (see Fig.~\ref{fig5}A). In this way, we obtained both a narrow short-term linewidth and an improved long-term stability of the laser (Fig.~\ref{fig5}B red traces) that is traceable to the SI definition of the second as a prerequisite for applications in precision spectroscopy. 

To illustrate the improved frequency measurement of the spectroscopy laser, the frequency of the laser was determined by two methods, i.e., first, by referencing its frequency to a local Global Positioning System disciplined (GPSD) Rb clock (SRS FS725 Rb standard disciplined to Symmetricom GPS-500), or second, by referencing it to the frequency disseminated from METAS. In either case, the frequency comb was optically locked to the \SI{729}{nm} laser.
In the first method, the frequency was given by
\begin{equation}\label{eq:Rb}
    f_{729}=2f_\textrm{CEO}+n_{729}f_\textrm{rep}+f_{729}^\textrm{beat}.
\end{equation}
Here, $f_\textrm{CEO}$ and $f_\textrm{rep}$ are the carrier-envelope-offset frequency (CEO) and the repetition-rate of the frequency comb, respectively, $n_{729}=1'644'167$ is the index of the frequency-comb tooth used to create a beat with the spectroscopy laser at frequency $f_{729}^\textrm{beat}$. All measured frequencies were referenced to the local GPSD Rb clock. The factor of 2 appears with $f_\textrm{CEO}$ because the spectrum of the comb is doubled in order to beat with the \SI{729}{nm} laser. The limiting factor in this method was the repetition-rate stability as measured by the GPSD Rb clock due to the large multiplication factor, $n_{729}$.
In the second method, the frequency was given by
\begin{equation}\label{eq:METAS}
    f_{729}=(2-x)f_\textrm{CEO}+x(f_{1572}-f_{1572}^\textrm{beat})+f_{729}^\textrm{beat}.
\end{equation}
Here, $f_{1572}$ is the frequency of the regeneration laser which was determined by METAS, $f_{1572}^\textrm{beat}$ is the beat frequency with the local frequency-comb tooth, $n_{1572}=762'813$, and $x=n_{729}/n_{1572}\approx2.16$. With this method, the local-frequency-comb repetition rate does not appear in the frequency comparison since we make a comparison between two optical frequencies. The limiting factor in this method is the stability of the disseminated frequency manifested in the link beat with the local frequency comb, $f_{1572}^\textrm{beat}$.

In Fig.~\ref{fig5}B, we show plots of the Allan deviation of the \SI{729}{nm} laser frequency obtained by both methods. In the first method (blue traces), the stability of the Rb clock limits the frequency determination to $\sim5\times10^{-13}$ at \SI{1000}{s}. The METAS standard delivered via the fiber link, however, provides more than two orders of magnitude improvement in the frequency traceability of the laser over the averaging time analyzed here (thick red trace). Already at an averaging time of \SI{1000}{s}, the statistical uncertainty of $\sim1\times 10^{-15}$ is better than the instability originating from the METAS standard compared to TAI. 

The Allan deviation of the link exhibits a plateau up to \SI{5}{s} integration time. This behavior is attributed to the time constant of the lock of the master laser to the SI-referenced frequency comb at METAS and the instability of the ULE cavity at METAS which has higher drifts and a lower finesse compared to the cavity used in Basel. 
The small shoulder at \SI{100}{s} is a reminiscence of the ULE cavity in Basel. The maximum at \SI{3000}{s} is attributed to the feedback loop of the slow-drift compensation in Basel. 

\subsection{Bit error rate measurement on the data traffic channels}
\begin{figure}[ht]
	\centering\includegraphics[width=0.48\textwidth,trim={0.4cm 0.4cm 0.1cm 0.3cm},clip]{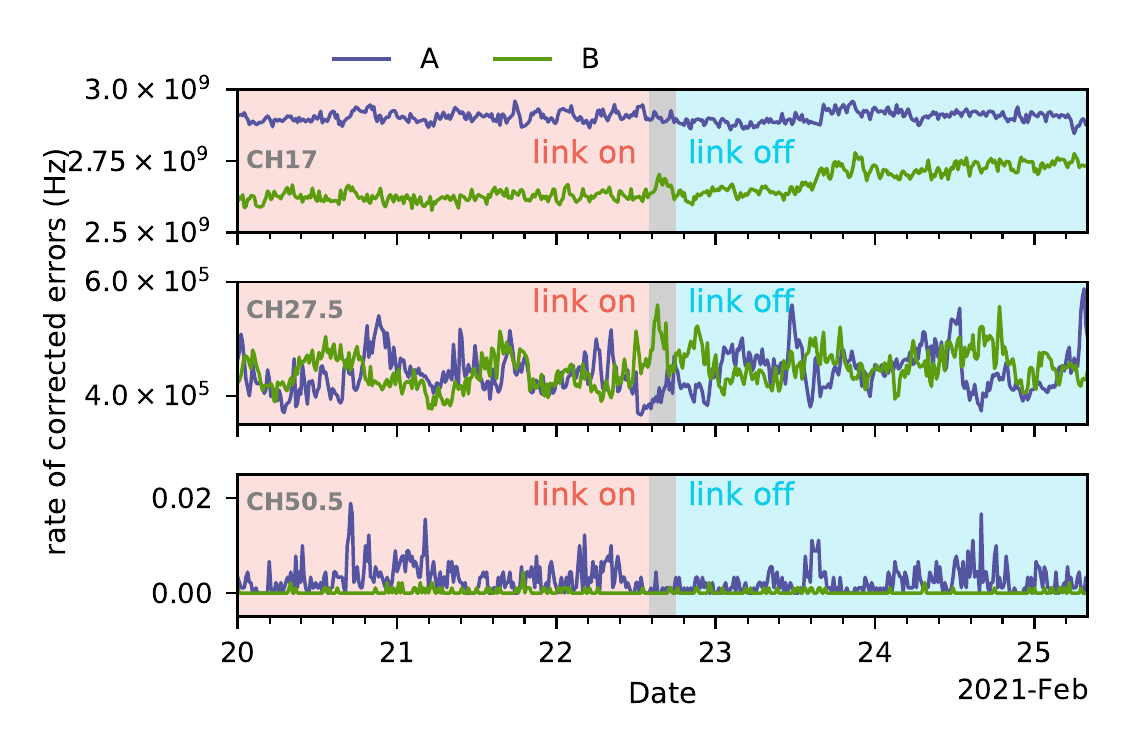}
	\caption{Rate of corrected errors between two transponders in 3 selected data channels (CH17, CH27.5 and CH50.5) sharing fibers with the metrological signal. Each channel was measured in both directions (lines A and B). In the red (blue) shaded area, the metrological signal was on (off). In the transitions time (gray area), the signal was partially on. The modulation schemes were \SI{200}{GBit/s} polarization 8 quadrature amplitude modulation (DP-8QAM) for CH17, \SI{100}{Gbit/s} dual polarization quadrature phase shift keying for CH27.5 and \SI{10}{Gbit} non-return-to-zero on-off keying for CH50.5.}
	\label{fig6}
\end{figure}

Multiplexing a new frequency channel into an existing data network requires careful evaluation of potential cross talk with the existing channels. In order to verify that our metrological signal in ITU-T CH07 does not disturb the existing data-communication channels in the \textit{C-band}, we exploited the integrated error detection and correction schemes (forward error correction) of the optical transponder systems in selected data traffic channels of the SWITCH network infrastructure. To this end, we observed the rate of corrected errors per second in the transponders over 2-3 days both in presence and absence of the metrological signal in the network, with the report interval of the transponders set to \SI{15}{min}. The measurement was performed on 14 different optical channels varying in fiber segments, frequency and data modulation schemes. All channels maintained error-free communication during the measurement times, demonstrating that all detected errors were corrected by the transponders. Time traces of the rate of corrected errors for 6 exemplary channels are shown in Fig.~\ref{fig6}. While there are clear day-to-day fluctuations, the data show no significant impact of our \SI{1572}{nm} signal on the number of corrected errors in the measured data channels for either of the 14 measurements. This observation validates our choice of a metrological \textit{L-band} frequency and confirms that optical dissemination in the \textit{L-band} does not affect telecommunication data transmission in the \textit{C-band}.

\section{Conclusion and outlook}
The results presented here validate the approach of multiplexing a channel for stabilized metrological-frequency dissemination located in the \textit{L-band} into an existing optical fiber data network. The performance of the network link allows for the dissemination of an SI-traceable frequency without significant degradation in stability and accuracy. Here we demonstrated the improvement of the frequency reference in a remote laboratory by two orders of magnitude compared to the previously employed GPSD Rubidium clock standard. As prospective applications of the present frequency-transfer network, we envisage precision spectroscopy of molecular ions \cite{germann_observation_2014, najafian20b} and of Rydberg atoms and molecules \cite{beyer_metrology_2018,peper_precision_2019}. From a metrology viewpoint, the network opens up the potential to compare atomic clocks, such as the primary frequency standard FoCS-2 at METAS \cite{jallageas_first_2018}, to other frequency standards. Furthermore, such a network can be used for frequency comparisons beyond the current stability given by primary frequency standards, such as optical clocks. The frequency-dissemination network presented here offers the advantage of being both cost effective and relatively simple in implementation in an existing fiber network, which is a prerequisite for the development of a  world-wide metrological network.

\begin{acknowledgments}
We thank A. Frank for support with electronics.
This project it funded by the Swiss National Science Foundation (SNSF), Sinergia grant nr. CRSII5\_183579. Additional funding from the SNSF through the National Centre of Competence in Research, Quantum Science and Technology (NCCR-QSIT) (SW) and through the grant nr. 743121 (FM) is acknowledged.
\end{acknowledgments}

\bibliography{sample}

\end{document}